\documentclass[preprintnumbers,amsmath,11pt,amssymb,floatfix,superscriptaddress,showpacs,nofootinbib]{revtex4}
\usepackage{graphicx}
\usepackage{epsfig}
\usepackage{bm}
\usepackage{amsfonts}

\begin{document}

\title{\Large A study on Ricci dark energy in bulk-brane interaction}

\author{Surajit Chattopadhyay}
\email{surajit_2008@yahoo.co.in, surajcha@iucaa.ernet.in}
\affiliation{ Pailan College of Management and Technology, Bengal
Pailan Park, Kolkata-700 104, India.}

\begin{abstract}
In this work we have investigated the effects of the interaction
between a brane universe and the bulk in which it is embedded.
Considering the effects of the interaction between a brane
universe and the bulk we have obtained the equation of state for
the interacting holographic Ricci dark energy density
$\rho_{\Lambda}=3c^2 (\dot{H}+2 H^2)$ in the flat universe. We
have investigated the impact of $c^2$ on the equation of state
$\omega_{\Lambda}$. Also, considering the power law for of the
scale factor, we have observed that non-trivial contributions of
dark energy which differ from the standard matter fields confined
to the brane is increasing with the evolution of the universe.
\end{abstract}


\maketitle

\section{Introduction}
The observed accelerated expansion of the universe is usually
attributed to the presence of an exotic kind of energy, called
``dark energy" (DE)\cite{DE1,DE2,DE3,DE4,DE5,DE6,DE7,DE88}.  The measurement of WMAP and the supernova analysis predicts that exactly $95.4915028\%$ of the cosmos must be dark energy \cite{add1,add2,add3}. A great
variety of DE models have been proposed so far. Models of dark energy are conveniently characterized by the equation-of-state parameter $\omega=p/\rho$, where $\rho$ is the energy density and $p$ is the pressure \cite{EoS1,EoS2,EoS3,EoS4}. The models are
discussed at length in the references mentioned above. Examples of
DE models are tachyon \cite{paddy1}, phantom \cite{cald},
Chaplygin gas \cite{chap}, K-essence \cite{kessence}, hessence
\cite{hessence}, dilaton \cite{dilaton}, holographic dark energy
\cite{HDE} etc. Observational evidences in support of DE are SNIa,
CMB and BAO \cite{DE8,DE9,DE10}. The Type Ia supernova (SNIa)
observations, cosmic microwave background (CMB) and baryon
acoustic oscillations (BAO) have confirmed that about $73\%$ of
the energy density of the present universe consists of dark
energy. The so called ``holographic dark energy" (HDE), based on
the holographic principle \cite{holoprinciple}, was proposed by Li
\cite{HDE1}. In holographic principle the black holes are regarded
as the maximally entropic objects of a given region and this
principle postulates that the maximum entropy inside this region
behaves non-extensively, growing only as its surface area
\cite{Ricci1}. Therefore, the number of independent degrees of
freedom is bounded by the surface area in Planck units. This
implies that an effective field theory with $UV$ cutoff $\Lambda$
in a box of size $L$ is not self consistent, if it does not
satisfy the Bekenstein entropy bound \cite{Ricci1}. It was
suggested suggested in \cite{Cohen} that the total energy in a
region of size $L$ should not exceed the mass of a black hole of
the same size. Based on this assumption, reference \cite{HDE1}
proposed HDE density as
\begin{equation}\label{HDE}
\rho_{h}=3c^2 M_{p}^2 L^{-2}
\end{equation}
where $c^2$ is a dimensionless constant. Hsu \cite{Hsu} took
Hubble horizon is taken as the $IR$ cut-off in HDE, and
subsequently the HDE is written as $\rho_{h}=3c^2 M_{p}^2 H^{2}$.
This consideration by \cite{Hsu} failed to give the accelerated
expansion of the universe. Gao et al \cite{Gaoetal} proposed a
holographic dark energy model in which the future event horizon is
replaced by the inverse of the Ricci scalar curvature, and dubbed
this model the Ricci dark energy model (RDE). The Ricci curvature
of FRW universe is given by
\begin{equation}\label{Ricci}
R=-6\left(\dot{H}+2H^2+\frac{k}{a^2}\right)
\end{equation}
The density of the RDE is therefore \cite{Xu}
\begin{equation}\label{RDEdensity}
 \rho_{\Lambda}=3c^2 M_{p}^2 R=3c^2 M_{p}^2\left(\dot{H}+2H^2+\frac{k}{a^2}\right)
\end{equation}
Cai et al. \cite{RDE2} pointed out that the Ricci dark energy can
be viewed as originated from taking the causal connection scale as
the $IR$ cutoff in the holographic setting. Here we mention some
significant works on RDE. Feng \cite{RDE3} investigated the
statefinder diagnostics of the RDE. Feng \cite{RDE4} investigated
the nature of RDE with respect to the holographic principle of
quantum gravity theory. Feng \cite{RDE5} regarded the $f(R)$
theory as an effective description for the acceleration of the
universe and reconstruct the function $f(R)$ from the RDE. Feng
and Li \cite{RDE6} investigated the viscous RDE model by assuming
that there is bulk viscosity in the linear barotropic fluid and
the RDE. In another work Feng and Zhang \cite{RDE7} discussed the
role of $c^2$ in the RDE density (equation \ref{RDEdensity}) and
revealed that if $c^2<0.5$ then equation of state of RDE will
evolve across the cosmological constant boundary $-1$ i.e. RDE
will behave like quintom. In the present work, we are going to
discuss the RDE in bulk-brane interaction.

In recent years, the idea that the universe is a brane embedded in
a higher-dimensional space, has got much attraction of the
researchers \cite{1,2,3,4,5,6,7,sarida1}. Friedmann equation on the brane
contains corrections to the usual four-dimensional equation
\cite{4}. Binetruy et al \cite{3} found a term $H\propto \rho$,
which is observationally problematic. The model is consistent if a
cosmological constant in the bulk and the tension on the brane are
considered. This leads to a cosmological version of the
Randall–Sundrum scenario of warped geometries \cite{4}. Bruck et
al \cite{4} considered an interaction between the bulk and the
brane, which is another non-trivial aspect of brane world
theories. The purpose of the present work is to disclose the
effect of the energy exchange between the brane and the bulk on
the evolution of the universe by considering the flow of energy
onto or away from the brane. Employing the RDE in a flat universe,
we obtain the equation of state parameter for RDE density. This
study is motivated by the works of \cite{setare1,sheykhi,Myung,kim}. In an interaction between the bulk and the brane,
\cite{setare1} considered holographic model of dark energy in
non-flat universe under the assumption that the CDM energy density
on the brane is conserved, but the holographic dark energy density
on the brane is not conserved due to brane–bulk energy exchange.
Sheykhi \cite{sheykhi} considered the agegraphic models of dark
energy in a braneworld scenario with branebulk energy exchange
under the assumption that the adiabatic equation for the dark
matter is satisfied while it is violated for the agegraphic dark
energy due to the energy exchange between the brane and the bulk.
In the study of \cite{sheykhi}, the author revealed the the
equation of state (EoS) parameter can evolve from the quintessence
to phantom. Myung and Kim \cite{Myung} introduced the brane-bulk interaction to discuss a limitation of the cosmological Cardy-
Verlinde formula which is useful for the holographic description of brane cosmology and showed that the
presence of the brane-bulk interaction, one cannot find the entropy representation of the
first Friedmann equation. Saridakis \cite{sarida1} considered a generalized version of holographic dark energy arguing that it must be considered in the maximally subspace of a cosmological model and showed that in the context of brane cosmology it leads to a bulk holographic dark energy which transfers its holographic nature to the effective $4D$ dark energy. Saridakis \cite{sarida2} applied the bulk holographic dark energy in general $5D$ two-brane models and extracted the Friedmann equation on the physical brane and showed that in the general moving-brane case the effective $4D$ holographic dark energy behaves as a quintom for a large parameter-space area of a simple solution subclass.

In the present study we consider an interaction
between the bulk and the brane, which is non-trivial aspect of
braneworld theories.  We shall discuss the flow of energy onto or
away from the brane-universe. Then, using the holographic RDE in
flat universe  obtained equation of state for interacting
RDE. Firstly, we review the formalism of bulk-brane energy
exchange. Then we apply this material to a brane-world cosmology
under the assumption that the cold dark matter energy density on
the brane is conserved, but the RDE density on the brane is not
conserved due to brane–bulk energy exchange.
\\

\section{Bulk-brane energy exchange}
The bulk-brane action is given by \cite{setare1,cai}
\begin{equation}\label{action}
S=\int d^5x\sqrt{-G}\left(\frac{R_5}{2\kappa_{5}^{2}-\Lambda_5+L_{B}^{m}}\right)+\int d^4x\sqrt{-g}(-\sigma+L_{b}^{m})
\end{equation}
where $R_5$ is the curvature scalar of the five-dimensional
metric, $\Lambda_5$ is the bulk cosmological constant and $\sigma$
is the brane tension, $L_{B}^{m}$ and $L_{b}^{m}$ are the matter
Lagrangian in the bulk and on the brane, respectively. We consider
the cosmological solution with a metric of the form
\cite{setare1,cai}
\begin{equation}\label{metric}
 ds^2=-n^2(t,y)dt^2+a^2(t,y)\gamma_{ij}dx^{i}dy^{j}+b^{2}(t,y)dy^{2}
\end{equation}
The non-zero components of Einstein tensor can be written as
\cite{setare1,cai}
\begin{equation}\label{g00}
  G_{00}=3\left[\frac{\dot{a}}{a}\left(\dot{a}{a}+\dot{b}{b}\right)-\frac{n^2}{b^2}\left(\frac{a''}{a}\frac{a'}{a}\left(\frac{a'}{a}-\frac{b'}{b}\right)\right)+k\frac{n^2}{b^2}\right]~,
\end{equation}
\begin{equation}\label{gij}
\begin{array}{c}
  G_{ij}=\frac{a^{2}}{b^{2}}\gamma_{ij}\left[\frac{a'}{a}\left(\frac{a'}{a}+2\frac{n'}{n}\right)-\frac{b'}{b}\left(\frac{n'}{n}+2\frac{a'}{a}\right)+2\frac{a''}{a}+\frac{n''}{n}\right]
  + \\
  \frac{a^{2}}{n^{2}}\gamma_{ij}\left[\frac{\dot{a}}{a}\left(-\frac{\dot{a}}{a}+2\frac{\dot{n}}{n}\right)-2\frac{\ddot{a}}{a}+\frac{\dot{b}}{b}\left(-2\frac{\dot{a}}{a}+\frac{\dot{n}}{n}\right)-\frac{\ddot{b}}{b}\right]-k\gamma_{ij}~, \end{array}
  \end{equation}

\begin{equation}\label{g05}
 G_{05}=3\left(\frac{n'}{n}\frac{\dot{a}}{a}+\frac{a'}{a}\frac{\dot{b}}{b}-\frac{\dot{a}'}{a}\right)~,
\end{equation}
\begin{equation}\label{g55}
  G_{55}=3\left[\frac{a'}{a}\left(\frac{a'}{a}+\frac{n'}{n}\right)-\frac{b^2}{n^2}\left(\frac{\dot{a}}{a}\left(\frac{\dot{a}}{a}-\frac{\dot{n}}{n}\right)+\frac{\ddot{a}}{a}\right)-k\frac{b^2}{a^2}\right]~,
\end{equation}
where $k$ denotes the curvature of space $k= 0, 1,−1$ for flat,
closed and open universe, respectively and $\gamma_{ij}$ is the
metric for the maximally symmetric three-dimensional space. The
primes and dots denote the derivatives with respect to $y$ and $t$
respectively. The three-dimensional brane is assumed at $y=0$. The
Einstein equations are $G_{\mu\nu}=\kappa_{5}^{2}T_{\mu\nu}$,
where the stress-energy momentum tensor has bulk and brane
components and can be written as \cite{setare1,cai}
\begin{equation}\label{stress}
  T^{\mu}_{\nu}=T^{\mu}_{\nu}|_{\sigma,b}+T^{\mu}_{\nu}|_{m,b}+T^{\mu}_{\nu}|_{\Lambda,B}+T^{\mu}_{\nu}|_{m,B}
\end{equation}
where,
\begin{equation}\label{sigmab}
  T^{\mu}_{\nu}|_{\sigma,b}=\frac{\delta(y)}{b}diag(-\sigma,-\sigma,-\sigma,-\sigma,0),
\end{equation}
\begin{equation}\label{lambdab}
  T^{\mu}_{\nu}|_{\Lambda,B}=diag(-\Lambda_{5},-\Lambda_{5},-\Lambda_{5},-\Lambda_{5},-\Lambda_{5}),
\end{equation}
\begin{equation}\label{mb}
  T^{\mu}_{\nu}|_{m,b}=\frac{\delta(y)}{b}diag(-\rho,p,p,p,0),
\end{equation}
where, $p$ and $\rho$ are the pressure and density on the brane, respectively. Integrating equations (\ref{g00}) and (\ref{gij}) with respect to $y$ around $y=0$ give the following jump conditions
\begin{equation}\label{jump1}
 a'_{+}=-a'_{-}=-\frac{\kappa_{5}^{2}}{6}a_{0}b_{0}(\sigma+\rho),
\end{equation}
\begin{equation}\label{jump2}
 n'_{+}=-n'_{-}=\frac{\kappa_{5}^{2}}{6}b_{0}n_{0}(-\sigma+2\rho+3p),
\end{equation}
Using equations (\ref{jump1}) and (\ref{jump2}), one can derive
\begin{equation}\label{field1}
  \dot{\rho}+3\frac{\dot{a}_{0}}{a_{0}}(\rho+p)=-\frac{2n_{0}^{2}}{b_{0}}T^{0}_{5},
\end{equation}
\begin{equation}\label{field2}
  \frac{1}{n_{0}^{2}}\left[\frac{\ddot{a}_{0}}{a_{0}}+\left(\frac{\dot{a}_{0}}{a_{0}}\right)^{2}-\frac{\dot{a}_{0}\dot{n}_{0}}{a_{0}n_{0}}\right]+\frac{k}{a_{0}^{2}}=\frac{\kappa_{5}^{2}}{3}\left(\Lambda_{5}+\frac{\kappa_{5}^{2}\sigma^{2}}{6}\right)\\
  -\frac{\kappa_{5}^{4}}{36}\left[\sigma(3p-\rho)+\rho(3p+\rho)\right]-\frac{\kappa_{5}^{2}}{3}T^{5}_{5},
\end{equation}
where, $T_{05}$ and $T_{55}$ are the $05$ and $55$ components of $T_{\mu\nu}|_{m,b}$ evaluated on brane. Assuming that the bulk matter relative to bulk vacuum energy is much less than the ratio of the brane matter to the brane vacuum energy, we can neglect the $T^{5}_{5}$ term and this can lead to the derivation of a solution that is largely independent of the bulk dynamics. Considering this approximation and concentrating on the low-energy region with $\rho/\sigma\ll 1$, equations (\ref{field1}) and (\ref{field2}) can be simplified into
\begin{equation}\label{field11}
 \dot{\rho}+3H(1+w)\rho=-2T^{0}_{5}=T,
\end{equation}
\begin{equation}\label{field22}
  H^{2}=\frac{8 \pi G_{4}}{3}(\rho+\chi)-\frac{k}{a^{2}}+\lambda,
\end{equation}
\begin{equation}\label{field33}
\dot{\chi}+4H\chi\approx 2T^{0}_{5}=-T.
\end{equation}
The auxiliary field $\chi$ incorporates non-trivial contributions
of dark energy which differ from the standard matter fields
confined to the brane. Hence, with the energy exchange $T$ between
the bulk and brane, the usual energy conservation is broken down.
\\

\section{RDE in the bulk-brane interaction}
In the present work, we are going to discuss Ricci dark energy
(RDE) in bulk-brane interaction. The bulk-brane interaction has
been studied for various aspects in the works of
\cite{setare1,setare2,cai}. We shall denote the energy density of
RDE by $\rho_{\Lambda}$. Since we shall consider two dark
components in the universe, namely, dark matter and dark energy,
we shall have $\rho=\rho_{\Lambda}+\rho_{m}$. We assume that the
adiabatic equation for the dark matter is satisfied, while it is
violated for the dark energy due to the energy exchange between
the brane and the bulk \cite{setare1,cai},
\begin{equation}\label{matter}
  \dot{\rho}_{m}+3H\rho_{m}=0,
\end{equation}
\begin{equation}\label{energy}
  \dot{\rho}_{\lambda}+3H(1+\omega_{\Lambda})\rho_{\Lambda}=T.
\end{equation}
The interaction between bulk and brane is given by the quantity
$T=\Gamma \rho_{\Lambda}$, where $\Gamma$ is the rate of
interaction. Considering $u=\frac{\chi}{\rho_{\Lambda}+\rho_{m}}$,
the above equations lead to \cite{setare1}
\begin{equation}\label{udot}
 \dot{u}=\left(\frac{3Hu\Omega_{\Lambda}}{\Omega_{\Lambda}+\Omega_{m}}\right)\left[\omega_{\Lambda}-\frac{1}{3}\left(\frac{\Omega_{m}}{\Omega_{\Lambda}}+1\right)-\frac{1+u}{u}\frac{\Gamma}{3H}\right]
\end{equation}
In our present work, we take $\lambda=0$ and $k=0$. Furthermore,
following \cite{setare1} we have chosen
\begin{equation}\label{gamma}
 \Gamma=3b^2 (1+u)H
\end{equation}

In flat $(k=0)$ FRW universe, the density of the RDE is given by
\cite{RDEFeng}
\begin{equation}\label{RDE}
  \rho_{\Lambda}=3c^2\left(\dot{H}+2H^2\right)
\end{equation}
Differentiating equation (\ref{field22}) with respect to $t$ and using equations (\ref{field11}) to (\ref{field33}) one has (considering $8\pi G_4=1$) for flat universe i.e. $k=0$
\begin{equation}\label{Hdot}
 \dot{H}=-\frac{1}{6}\left[3\rho_{\Lambda}(1+\omega_{\Lambda})+3\rho_m+4\chi\right]
\end{equation}
Using equations (\ref{Hdot}) and (\ref{field22}) in equation
(\ref{RDE}) we get
\begin{equation}\label{Omegad}
    \omega_{\Lambda}=\frac{1}{3}\left(1+\frac{\Omega_{m}}{\Omega_{\Lambda}}\right)-\frac{2}{3c^2}
\end{equation}
 where, it is defined
\begin{equation}\label{Fractional}
 \Omega_{\Lambda}=\frac{\rho_{\Lambda}}{3H^2},~~\Omega_{m}=\frac{\rho_{m}}{3H^2},~~\Omega_{\chi}=\frac{\chi}{3H^2}.
\end{equation}
and subsequently
\begin{equation}\label{Fracsum}
   u=\frac{1-\Omega_{\Lambda}-\Omega_{m}}{\Omega_{\Lambda}}
\end{equation}
Using $\Omega_{m0}=0.27$ and $\Omega_{\Lambda0}=0.73$ we plot the
$\omega_{\Lambda}$ in figure 1. In this figure we observe that the
EoS parameter is evolving from $<-1$ to $>-1$ as we are changing
the value of $c^2$. This implies a transition from quintessence to
phantom. Moreover, we find that for $c^2<0.5$ the EoS is staying
below $-1$ and for $c^2>0.5$ the EoS is staying above $-1$. In our
calculation, we have taken $\Gamma=3 b^2(1+u)H$ Furthermore, we
have considered scale factor in power law form.

\begin{figure}[h]
\includegraphics[width=16pc]{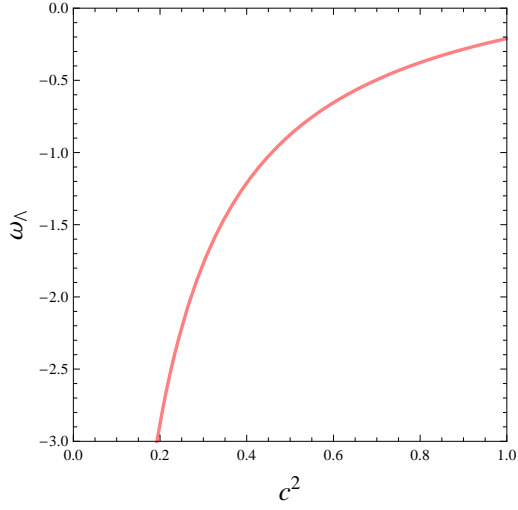}
\caption{\label{label}Variation of $\omega_{\Lambda}$ with $c^2$ pertaining to RDE. We have taken $\Omega_{\Lambda0}=0.73$ and $\Omega_{m0}=0.27$.}
\end{figure}

From equation (\ref{Omegad}) we can
write
\begin{equation}\label{uasa}
 \dot{u}=-\left(\frac{3Hu\Omega_{\Lambda}}{\Omega_{\Lambda}+\Omega_{m}}\right)\left[\frac{2}{3c^2}+\frac{b^2(1+u)^2}{u}\right]
\end{equation}
Since we have $\Omega_{\Lambda}+\Omega_{m}=(1+u)^{-1}$ the
equation (\ref{uasa}) can be rewritten as
\begin{equation}\label{uasanew}
   \dot{u}=\frac{3Hu(1+u)}{\Omega_{\Lambda}}\left[\frac{2}{3c^2}+\frac{b^2(1+u)^2}{u}\right]
\end{equation}
Let us select the scale factor $a$ in the power law form i.e.
$a(t)=a_{0}t^{n}$. Then
\begin{equation}\label{powerlaw}
   H=\frac{n}{t}
\end{equation}
Subsequently, equation (\ref{uasanew}) becomes
\begin{equation}\label{differntialeqn}
\dot{u}=-\frac{3n^2u(1+u)}{tc^2
(2n-1)}\left[\frac{2}{3c^2}+\frac{b^2(1+u)^2}{u}\right]
\end{equation}
\begin{figure}[h]
\includegraphics[width=16pc]{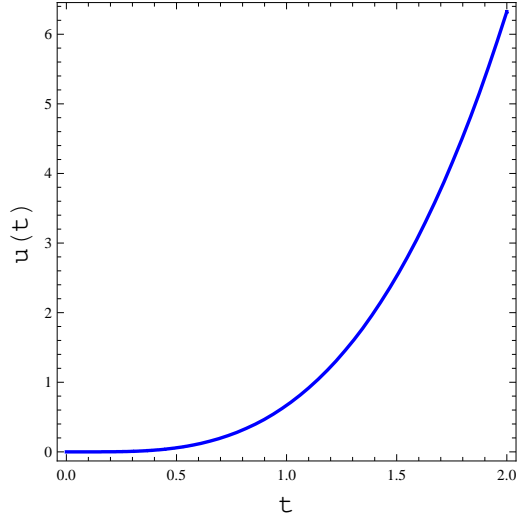}
\caption{\label{label}Pattern of $u$ with evolution of the universe under power law expansion. We have taken $c=0.7$,$n=1/3$ and $b=2$.}
\end{figure}
In figure 2 we have plotted $u$ against $t$. We observe that $u$
is increasing with time. This indicates that non-trivial
contributions of dark energy which differ from the standard matter
fields confined to the brane is increasing with the evolution of
the universe.
\\
\section{Conclusion}
In this work we have investigated the effects of the interaction
between a brane universe and the bulk in which it is embedded.
Considering the effects of the interaction between a brane
universe and the bulk we have obtained the equation of state for
the interacting holographic Ricci dark energy density
$\rho_{\Lambda}=3c^2 (\dot{H}+2 H^2)$ in the flat universe. We
have seen that there is a crossing of the barrier
$\omega_{\Lambda}=-1$ as $c^2$ shifts from $<0.5$ to $>0.5$. If
$c^2<0.5$ the EoS behaves like phantom $(\leq -1)$ and if
$c^2>0.5$ the EoS behaves like quintessence $(\geq -1)$. Next we
have chosen a particular form of scale factor $a(t)=a_{0}t^{n}$.
We expressed $\dot{u}$ as a function of $t$ and solving
numerically we plotted $u$ against $t$ and we found an increasing
behavior in $u$. We observed that non-trivial contributions of
dark energy which differ from the standard matter fields confined
to the brane is increasing with the evolution of the universe.
\\
\textbf{Acknowledgement}\\
 The author wishes to acknowledge the
financial support from the Department of Science and Technology,
Govt. of India under the Fast Track Scheme for Young Scientists.
The Grant No. is SR/FTP/PS-167/2011.\\

\end{document}